\documentclass[twocolumn,showpacs,preprintnumbers,amsmath,amssymb]{revtex4}

\usepackage{graphicx}
\usepackage{dcolumn}
\usepackage{bm}

\begin{document}

\title{Low-field microwave absorption in  epitaxial La$_{0.7}$Sr$_{0.3}$MnO$_{3}$ films resulting from the angle-tuned ferromagnetic resonance in the multidomain state }
\author{M. Golosovsky\footnote{e-mail michael.golosovsky@mail.huji.ac.il,
Permanent address: the Racah Institute of Physics, the Hebrew University of Jerusalem, 91904 Jerusalem, Israel},
 P. Monod,}%
\affiliation{Laboratoire de Physique du Solide, ESPCI, 10 rue Vauquelin, 75231 Paris Cedex 05, France}
\author{P.K. Muduli and R.C. Budhani}%
\affiliation{Department of Physics,Indian Institute of Technology,Kanpur, 208016, India\\}
\date{\today} 

\begin{abstract}
We studied magnetic-field induced microwave absorption  in 100-200 nm thick La$_{0.7}$Sr$_{0.3}$MnO$_{3}$ films on SrTiO$_{3}$ substrate and found a low-field absorption with a very peculiar angular dependence: it appears only in the oblique field and is absent both in the parallel and in the perpendicular  orientations. We demonstrate that this low-field absorption results from the ferromagnetic resonance in the multidomain state (domain-mode resonance). Its unusual angular dependence arises from the interplay between the parallel  component of the  magnetic field that drives the film into multidomain state and  the perpendicular field component that controls the domain width through its effect on domain wall energy. The low-field microwave absorption in the multidomain state can be a tool to probe domain structure in magnetic films with in-plane magnetization.  
\end{abstract}

\maketitle

\section{introduction}

Low-field microwave absorption  is a useful tool of detection of magnetic transitions in bulk materials and thin films \cite{Haddon,Bohandy,Owens}. It is also known as nonresonant, zero-field, or magnetically-modulated microwave absorption  and it was used to find the traces of ferromagnetic and superconducting phases in bulk  materials \cite{Haddon,Veinger}. The low-field  absorption in magnetic materials has multiple sources that operate in different frequency ranges \cite{Dionne}. The dominant contribution at radiofrequencies arises from magnetoimpedance \cite{Valenzuela,Montiel,Stanescu} (the field dependence of the skin-depth) and domain wall resonances \cite{Bahlmann,Synogach,Vukadinovic2000}, while at microwave frequencies it is commonly attributed to various  ferromagnetic resonances in the magnetically unsaturated state.  These include non-aligned FMR mode \cite{Prinz,Baberschke}, the natural FMR \cite{Lofland}, off-resonant absorption associated with the FMR tail \cite{Lee,Rivoire}, and the domain mode resonances \cite{Artman,Hasty,Salansky,Buznikov,Layadi,Vukadinovic1995,Ebels,Bi,Acher}. Magnetoresistive absorption in conducting ferromagnetics  contributes at all frequencies and is most spectacular at the field when magnetization changes abruptly, namely at zero field \cite{Zakhidov,Dubowik} and  upon transition from the unsaturated to saturated state  \cite{Fukui,Krebs,Vittoria,Belyaev,Demidov}. 

Manganites are conducting ferromagnetics with exceptionally high magnetoresistance \cite{Salamon}. Their microwave properties in magnetic field  have been extensively studied and the low-field microwave absorption was found in many compounds. While the dominant contribution in bulk manganites arises from magnetoimpedance \cite{Lofland,Budak}   there are many observations of the low-field microwave absorption in  thin films that can't be explained in this way. Some of these observations were attributed  to magnetoresistance in the magnetically-unsaturated state \cite{Stanescu,Lyfar,Golos} while the origin of others \cite{Musa,Aswal,Demidov,Lofland} is unclear. 

Here, we studied microwave absorption in high-quality epitaxial La$_{0.7}$Sr$_{0.3}$MnO$_{3}$ (LSMO) films  in the presence of magnetic field and observed the low-field microwave absorption peak with a very peculiar angular dependence: it appears only in the oblique field. We show that this peak is not related to magnetoresistance but arises from the ferromagnetic resonance in the multidomain state. 

\section{Domain mode resonance in a thin film with "easy-plane" anisotropy in the oblique magnetic field}
The theory of the ferromagnetic resonance in the multidomain state  was originally developed in relation to bubble or stripe domains in ferromagnetic films  with  out-of-plane anisotropy \cite{Artman,Layadi,Buznikov,Salansky}. The few studies of films with in-plane magnetization were restricted to the case of uniaxial  anisotropy  and parallel field orientation \cite{Hasty,Salansky}. Here, we consider a ferromagnetic film with  biaxial  easy-plane anisotropy in oblique field and calculate its high-frequency susceptibility following the approach of  Refs. \cite{Hasty,Buznikov,Salansky}.
 
\subsection{Magnetostatics}
\begin{figure}[ht]
\includegraphics[width=0.55\textwidth]{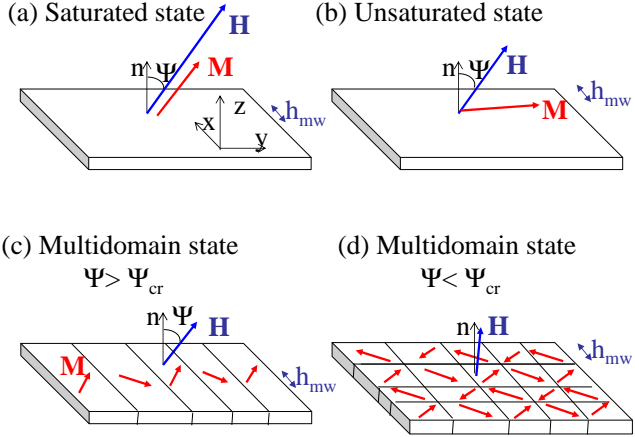}
\caption{Schematic drawing of a thin ferromagnetic film with biaxial easy-plane anisotropy in magnetic field $\mathbf{H}$ that makes angle $\Psi$ with the normal to the film  $\mathbf{n}$.  $\mathbf{M}$  is magnetization, $H_{\perp},H_{\parallel}$ are the perpendicular and parallel anisotropy fields, hard axes are along the  $x-$ and $y-$  directions. (a) Saturated state. Magnetization is collinear with the field. (b) Unsaturated state. Magnetization is not collinear with the field and tends to lie in the film plane. (c) Multidomain state, $\Psi>\Psi_{cr}$.  The film is split into parallel domains with the preferential orientation of the walls in  $x$-direction (i.e. perpendicular to $H_{y}$).  In the presence of a small microwave magnetic field $\mathbf{h}_{mw}$, which is  perpendicular to the $\mathbf{H}$ field  and has a component parallel to domain walls, the domain mode resonance can be excited. (d) Multidomain state, $\Psi<\Psi_{cr}$. The resulting domain structure is not clear and we plot only one of the possibilities. The film is split into irregular domains without preferential orientation of the walls, in such a way that the domain mode resonance is impeded.
}
\label{fig:scheme}
\end{figure}

Consider a thin ferromagnetic film  in oblique magnetic field $H$ (Fig.\ref{fig:scheme}). Its  free energy is
\begin{equation}
W=W_{\parallel}+W_{\perp}+W_{domain}+W_{Zeeman}+W(M)
\label{energy-1}
\end{equation}
where $W_{\parallel}$ and  $W_{\perp}$ are the in-plane and out-of-plane anisotropy energies, $W_{domain}$ is the energy associated with domain walls and stray field of domains, $W_{Zeeman}$ is the Zeeman energy, and  $W(M)$ absorbs all contributions that do not depend on field  orientation.  We assume a strong  easy-plane  anisotropy, $W_{\perp}=\frac{N_{z}M_{z}^2}{2}$, and a weak in-plane anisotropy, $W_{\parallel}=\beta\frac{M_{x}^4+M_{y}^4}{4M^{2}}$. Here, $M$ is the saturation magnetization, $N_z$ is the effective demagnetizing factor that includes both the shape and the crystalline anisotropies, $x,y$ are hard  axes  and $\beta$ is the (positive) in-plane anisotropy constant, whereas $\beta <<N_{z}$.  We assume that the external field has only $y-$ and $z-$ components, in such a way that the field projection on the film plane is parallel to one of the in-plane hard axes.

Figure \ref{fig:scheme} schematically shows the magnetic structure of the film when the external field slowly decreases to zero. At high  field the film is in the single domain state, the magnetization lies in the $y-z$ plane (Fig.\ref{fig:scheme}a,b), and $M_{x}=0$. At lower field the magnetization deviates towards the in-plane easy axes (which lie at 45$^0$ to $x-$ and $y$-axes) and $M_{x}\neq 0$. The film splits onto parallel domains with  domain walls in the $x-z$ plane (Fig.\ref{fig:scheme}c). When the field is nearly perpendicular to the film, the resulting magnetic structure is not clear. We believe that the film exhibits an irregular domain pattern (Fig.\ref{fig:scheme}d). 

Consider the structure shown in Fig.\ref{fig:scheme}c. Magnetizations of the adjacent domains, $\mathbf{M}_{1},\mathbf{M}_{2}$, have the same magnitude but different orientation. The free energy of a domain is:
\begin{eqnarray}
W_{1}=\beta\frac{M_{1x}^4+M_{1y}^4}{4M^{2}} -(M_{1y}H_{y}+M_{1z}H_{z})+ \nonumber\\ \frac{N_{z}M_{1z}^2}{2}+\frac{N_{y}(M_{1y}-M_{2y})^2}{4}+\frac{\lambda M^2}{2} 
\label{energy}
\end{eqnarray}
where  $N_y,N_{z}$ are the demagnetizing factors  in the $y-$ and $z-$ directions,  $M^2= M_{1x}^2+M_{1y}^2+M_{1z}^2$ is the magnetization,  $H_{y}=H\sin{\Psi},H_{z}=H\cos{\Psi}$ are projections of the field onto $y-$ and $z-$axes, and $\lambda$ is the Lagrange multiplier. The contribution of the domain wall energy and stray field of  domains was neglected. The effective field is
\begin{eqnarray}
\mathbf{H}^{eff}_1=-\frac{\partial W_1}{\partial \mathbf{M_{1}}}=-\mathbf{i} M_{1x}\left(\lambda +\frac{\beta M_{1x}^{2}}{M^2}\right)+\nonumber\\
\mathbf{j}\left[H_{y}-\lambda M_{1y}+\frac{\beta M_{1y}^{3}}{M^2}-\frac{N_{y}}{2}(M_{1y}-M_{2y})\right]\nonumber\\
+\mathbf{k}[H_{z}-(\lambda +N_{z}) M_{1z}]
\label{field1}
\end{eqnarray}
A similar expression holds for $\mathbf{H}^{eff}_2$.  The equilibrium condition,  $\mathbf{H}^{eff}_{1}=\mathbf{H}^{eff}_{2}=0$, yields  magnetization components $M_{1y}^{eq}=M_{2y}^{eq},M_{1x}^{eq}=-M_{2x}^{eq}$ and 
\begin{eqnarray}
M_{x}^{eq}\left[\lambda+\beta\left(\frac{M_{x}^{eq}}{M}\right)^{2}\right]=0;\nonumber\\
M_{y}^{eq}\left[\lambda+\beta \left(\frac{M_{y}^{eq}}{M}\right)^{2}\right]=H_{y};\nonumber\\
M_{z}^{eq}(\lambda+N_{z})=H_{z}.
\label{biaxial}
\end{eqnarray}
where indices 1,2 were dropped.  The single domain state corresponds to  $M_{x}^{eq}=0,\lambda\neq 0$, while the multidomain state corresponds to finite $M_{x}$ and $\lambda=-\beta (M_{x}^{eq}/M)^{2}$.  The onset of the multidomain state occurs when $M_{x}^{eq}=0$ and $\lambda=0$. Equation \ref{biaxial} yields the  field $H_{0}$ that corresponds to the onset of the domain state,
\begin{eqnarray}
\left(\frac{H_{0}\sin{\Psi}}{H_{\parallel}}\right)^{\frac{2}{3}}+\left(\frac{H_{0}\cos{\Psi}}{H_{\perp}}\right)^2=1
\label{split}
\end{eqnarray}
where  $H_{\parallel}=\beta M$ and $H_{\perp}=N_{z}M$ are the parallel and the perpendicular  anisotropy fields. For weak in-plane anisotropy, $H_{\parallel}<<H_{\perp}$, Eq.\ref{split} yields $H_{0}\approx H_{\parallel}/\sin{\Psi}$ for $\Psi>>\Psi_{cr}$, and $H_{0}\approx H_{\perp}$ for $\Psi<<\Psi_{cr}$ where $\Psi_{cr}=H_{\parallel}/H_{\perp}$ is the critical angle that delineates between two regimes that exhibit different domain patterns. 
  \begin{itemize}
    \item Oblique field, $\Psi>\Psi_{cr}$. Upon decreasing field the film goes from the magnetically-saturated  to the unsaturated state when the following condition is met: $H\cos{\Psi}=H_{z}\approx H_{\perp}$. At lower field when  $H\sin{\Psi}=H_{y}=H_{\parallel}$ the  film goes into multidomain state and splits into parallel domains. The preferential direction of the domain walls is determined by the field projection  onto the film plane (Fig.\ref{fig:scheme},c).
    \item Nearly perpendicular orientation, $\Psi<\Psi_{cr}$.  When the film  goes from the magnetically-saturated  to  unsaturated state at $H\cos{\Psi}=H_{z}\approx H_{\perp}$,  the second condition, $H_{y}=H\sin{\Psi}\approx H_{\parallel}$, is already met. The in-plane component of the external field at the onset of the domain state is too small to impose preferential orientation of the domain walls, hence the film most probably splits into  irregular domain pattern (Fig.\ref{fig:scheme}d). 
\end{itemize}     

\subsection{Domain mode resonance}
To derive the dynamic response of the film with parallel domains to the microwave magnetic field  $\mathbf{h}=he^{iwt}$ oriented along the  domain walls (Fig.\ref{fig:scheme}c) we use the Landau-Lifshitz equation without damping
\begin{equation}
\frac{\partial \mathbf{M}}{\partial t}=-\gamma [\mathbf{M}\times\mathbf{H}^{eff}]
\label {dynamics}
\end{equation}
where $\gamma$ is the gyromagnetic ratio. We consider the microwave field as a small perturbation that induces rotation of magnetization of  each domain, in other words we assume  $\mathbf{M}=\mathbf{M}^{eq}+\mathbf{m}$ where $\mathbf{M}^{eq}$ is the static equilibrium magnetization and $\mathbf{m}$ is a small time-dependent contribution. The effective field also acquires small dynamic contribution, $\delta H^{eff}=\frac{\partial {H^{eff}}}{\partial {\mathbf{M}}} \mathbf{m}+\mathbf{h}_{mw}$, where $H^{eff}$ is given by Eq.\ref{field1}.  We  linearize Eqs.\ref{field1},\ref{dynamics} and find the dynamic effective  field:
\begin{eqnarray}
\mathbf{\delta H}^{eff}_1=\mathbf{i} (h-\lambda m_{1x})- 
\mathbf{j}[(\beta+\lambda) m_{1y}+\nonumber\\\frac{N_{y}}{2}(m_{1y}-m_{2y})]-
\mathbf{k}(\lambda +N_{z}) m_{1z}
\label{deltaH}
\end{eqnarray}
A similar expression holds for domain 2. Solution of Eqs. \ref{dynamics},\ref{deltaH} yields two eigenfrequencies
\begin{widetext}
\begin{eqnarray}
\omega_{l}^{2}=\gamma^{2}M_{x}^{2}\beta\left(N_{z}\frac{5M_{y}^2- M_{x}^2}{M^{2}}+\beta \frac{6M_{z}^{2}M_{y}^{2}- 2M_{z}^{2}M_{x}^{2}-5M_{y}^{2}M_{x}^{2}+M_{x}^{4}} {M^{4}}\right)
\label{eigenfrequency1a}
\end{eqnarray}
\begin{eqnarray}
\omega_{h}^{2}=\gamma^{2}M_{x}^{2}\left[N_{y}N_{z}+\beta\left( N_{y}\frac{4M_{z}^{2}-M_{x}^{2}}{M^2}+N_{z}\frac{5M_{y}^2- M_{x}^2}{M^{2}}+\beta \frac{6M_{z}^{2}M_{y}^{2}- 2M_{z}^{2}M_{x}^{2}-5M_{y}^{2}M_{x}^{2}+M_{x}^{4}} {M^{4}}\right)\right]
\label{eigenfrequency2a}
\end{eqnarray}
\end{widetext}
Since  magnetization in the unsaturated state is almost parallel to the film, we can simplify these cumbersome expressions by omitting the terms with $M_{z}$.  Equations \ref{eigenfrequency1a}, \ref{eigenfrequency2a} reduce to:
\begin{equation}
\omega_{l}^{2}\approx\gamma^{2}M_{x}^{2}\beta \left(N_{z}-\frac{\beta M_{x}^{2}}{M^2}\right)\left(\frac{5M_{y}^2- M_{x}^2}{M^{2}}\right)
\label{eigenfrequency1}
\end{equation}
\begin{equation}
\omega_{h}^{2}\approx\gamma^{2}M_{x}^{2}N_{y}\left(N_{z}-\frac{\beta M_{x}^{2}}{M^2}\right)
\label{eigenfrequency2}
\end{equation}
Both eigenfrequencies depend on external field (through $M_{x}$) and go to zero at the onset of the domain state where $M_{x}=0$.  The lower frequency, $\omega_{l}$, almost does not depend on the domain structure since in this  "acoustic"  mode the magnetizations of neighboring domains precess almost in phase. On the other hand, the higher frequency, $\omega_{h}$, depends on the domain shape (through the demagnetizing factor $N_{y}$) since in this  "optical" mode the  magnetizations of neighboring domains precess in antiphase and the resulting magnetization has poles on the domain walls. The complex susceptibility 
\begin{equation}
\chi_{xx}=\frac{m_{1x}+m_{2x}}{2h}\approx\frac{\gamma^{2}M_{y}^{2}\left(N_{z}-\frac{\beta M_{x}^{2}}{M^{2}}\right)}{\omega_{h}^2-\omega^2}
\label{chi}
\end{equation}
exhibits resonance at $\omega=\omega_{h}$. The resonant frequency  can be crudely estimated from  Eq.\ref{eigenfrequency2}, 
\begin{equation}
\omega_{h}\sim \gamma M_{x}\sqrt{N_{y}N_{z}}.
\label{eigenfrequency3}
\end{equation}
Since the domain demagnetizing factors are $N_{z}\approx\frac{w}{d+w},N_{y}\approx\frac{d}{d+w}$ where $d$ is the film thickness and $w$ is the domain width \cite{Hasty},   Eq.\ref{eigenfrequency3} yields $\omega_{h}\approx \gamma M_{x}\frac{\sqrt{wd}}{w+d}$ (we neglected here the contribution of stress and crystallographic anisotropy to $N_{z}$).  When the field is decreased to zero the magnetization in domains rotate (affecting $M_{x}$), domain width $w$ also changes, in such a way that $\omega_{h}$ and, correspondingly, the microwave susceptibility (Eq.\ref{chi}) can pass through the resonance. For nearly perpendicular field orientation, when the film  splits into irregular domains (Fig.\ref{fig:scheme}d) such resonance is impeded.

Although the above model assumes that the field projection on the film plane is exactly along the hard magnetization axis (Fig.\ref{fig:scheme}) this assumption is probably not too restrictive. To observe the domain mode resonance it is necessary to have a system of parallel domains with alternating magnetizations and the component of the microwave magnetic field parallel to the domain walls. This requirement can be satisfied  when the  field projection on the film considerably deviates from the hard magnetization axis, especially for the films with biaxial in-plane anisotropy. 

\begin{figure}[ht]
\includegraphics[width=0.4\textwidth]{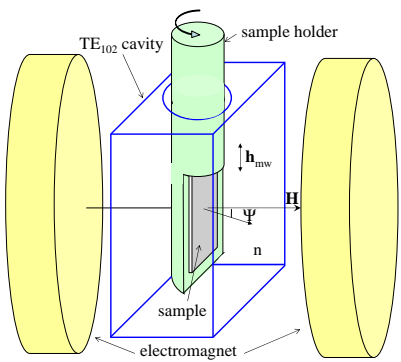}
\caption{Schematic drawing of the sample in magnetic field. A  sample holder with attached film on substrate is mounted  in the center of  the 9.4 GHz $TE_{102}$ resonant cavity. The polar angle of the magnetic field  $\mathbf{H}$ is varied by sample rotation. The microwave magnetic field $\mathbf{h}_{mw}$ is always perpendicular to $\textbf{H}$  and  parallel to the film plane.
}
\label{fig:setup}
\end{figure}

\section{Experiment and comparison to model}
\subsection{Experimental procedure}
To measure the low-field microwave absorption associated with the domain mode resonance we used a bipolar $X$-band  Bruker ESR spectrometer, a  $TE_{102}$ resonant cavity, and an Oxford helium flow  cryostat (Fig.\ref{fig:setup}). 
We studied   La$_{0.7}$Sr$_{0.3}$MnO$_{3}$  films with thicknesses $d=$50, 100, 150, and 200 nm. The films  were grown by the pulsed laser deposition technique on the $\sim 5\times 5\times 1$ mm$^3$  (001) SrTiO$_{3}$ substrates in two different laboratories \cite{Budhani,Mechin} and have $T_{C}$ of 330 K. The FMR measurements in the parallel field showed biaxial (four-fold) in-plane anisotropy of all these epitaxial films.  We measured magnetically-modulated microwave absorption and dispersion in  $1\times1$ mm$^2$ pieces of these films when the magnetic field was swept  from large negative to large positive values (typically, from -10 kOe to 10 kOe). To change the  field orientation with respect to the film we rotated the sample whereas the microwave magnetic field was always parallel to the film plane and perpendicular to the dc magnetic field. To find the exact perpendicular orientation we slightly tilted the sample holder in different directions, measured the field of the ferromagnetic resonance (FMR), and found the orientation in which the resonant field attained its maximum value. In what follows we focus on the results for the different pieces of the 200 nm thick film. We got similar results  for the 100 nm and 150 nm thick films but not for the 50 nm thick film. According to our interpretation this difference can be related to the different domain structure in thin films  whose thickness is comparable to the domain wall width.

\begin{figure}[ht]
\includegraphics[width=0.4\textwidth]{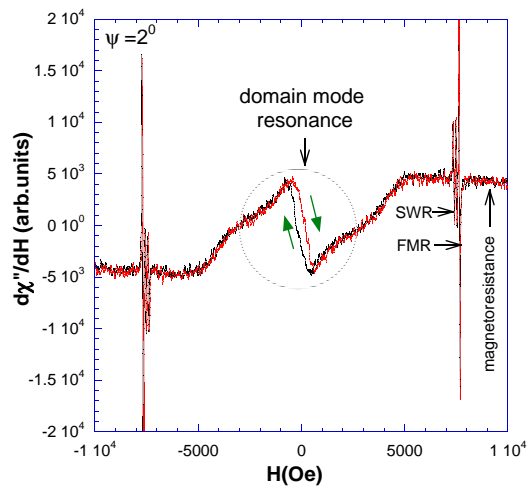}
\caption{Derivative microwave absorption for a 200 nm thick LSMO film on (001) STO substrate in magnetic field deviating  from the perpendicular orientation  by $\Psi=2^0$. The ferromagnetic (FMR) and the spin-wave resonances (SWR) appear at $\pm$ 7.6 kOe, the absorption baseline results from magnetoresistance. Note strong zero-field feature between -2 kOe and 2 kOe that shows considerable hysteresis. We attribute it to the domain mode resonance.
}
\label{fig:der-abs}
\end{figure}

\subsection{Low-field microwave absorption at 295 K}
Figure \ref{fig:der-abs} shows  derivative microwave  absorption  in oblique field for the 200 nm thick film and for two opposite directions of the field sweep.  The narrow peaks  at $\pm$ 7.6 kOe arise from the ferromagnetic and spin-wave resonances, the baseline arises from magnetoresistance, the pronounced zero-field feature exhibiting hysteresis is attributed to domain mode resonance. This zero-field feature does not depend  on the magnitude (1-10 Oe) or frequency of the modulation field (1-100 kHz).

\begin{figure}[ht]
\includegraphics[width=0.5\textwidth]{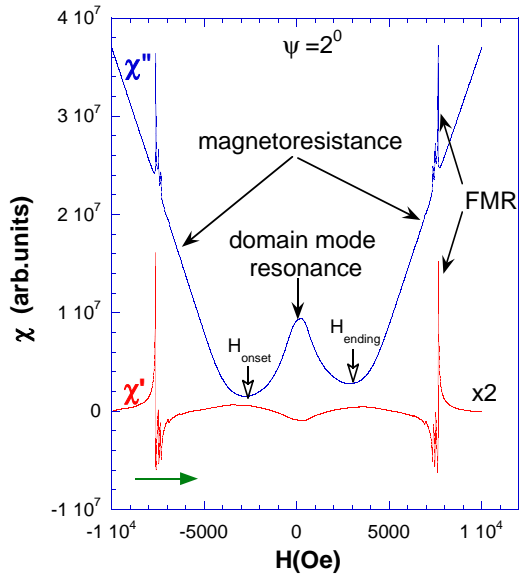}
\caption{Integrated microwave absorption $\chi^{''}$ and dispersion $\chi^{'}$for the  film shown in Fig.\ref{fig:der-abs}. The linear  baseline in the absorption spectrum results from  magnetoresistance while the zero-field  peak and the corresponding zero-field dip in the dispersion spectrum  are attributed to the domain mode resonance. $H_{onset}$ and $H_{ending}$ indicate the field range where the low-field absorption appears. 
}
\label{fig:int-abs}
\end{figure}

\begin{figure}[ht]
\includegraphics[width=0.5\textwidth]{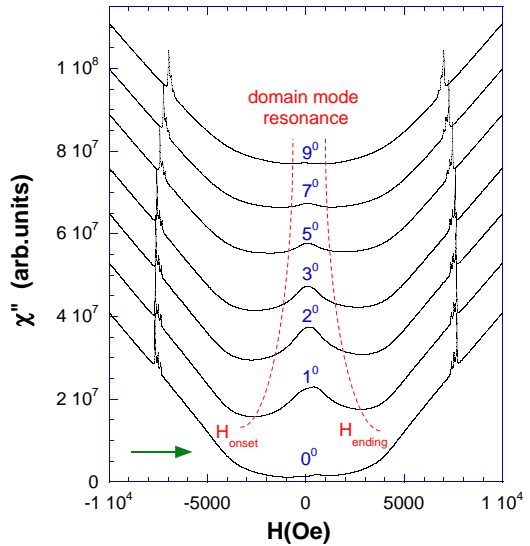}
\caption{Integrated microwave absorption  at different deviations of the field from the perpendicular orientation. The low-field absorption peak appears in the narrow angular range of $8^{0}>\
|\Psi|>1^{0}$ close to the perpendicular orientation. The red dashed line shows the field interval where this low-field absorption appears. This interval narrows  with increasing $\Psi$. The field corresponding to the maximum absorption also shifts with the angle and depends on the direction of the field sweep. The curves are vertically shifted for clarity.
}
\label{fig:int-abs1}
\end{figure}

Figure \ref{fig:int-abs} shows corresponding integrated   absorption $\chi''$ and dispersion $\chi'$  (details of the integration procedure are in Ref. \cite{Golos}). The  absorption baseline linearly increasing with field arises from magnetoresistance  (rigorously speaking, from magnetoconductance) and is absent in the dispersion spectrum, as expected. The focus of our study is a  low-field peak in the absorption spectrum and a corresponding  dip  in the  dispersion spectrum. We attribute both these features to the tail of the domain mode resonance. Indeed, we substitute into Eq.\ref{eigenfrequency3} $w=$0.5 $\mu$m \cite{Houwman,Lloyd},  $d=200$ nm,  $4\pi M_{x}=4\pi M=$ 4 kG and find $\omega_{h}/2\pi=$5.1 GHz. This value is below but not far away from our operating frequency of 9.4 GHz, hence the tail of the  resonance should introduce positive contribution to $\chi''$ and negative contribution to $\chi'$ (Eq.\ref{chi}, $\omega_{h}<\omega$)  that conforms to our observations (Fig.\ref{fig:int-abs}).

Note that the magnitude of  the zero-field absorption peak is close to the magnitude of the FMR peak.  Since the FMR absorption peak arises from the microwave absorption in the whole film, the zero-field absorption peak is an intrinsic effect (if it were related to the microwave absorption associated with some defect, its magnitude should be much smaller). Moreover, when we took a film  with the size $5\times 5$ mm$^2$ where and cut it to several smaller pieces we observed the zero-field absorption peak in each piece.

\subsubsection{Angular dependence}
Figure \ref{fig:int-abs1} shows that the low-field absorption peak appears in the oblique field and disappears when the field deviation from the perpendicular orientation is too large or too small. Since the conventional protocol of the angular-dependent microwave absorption measurements with the ESR spectrometer starts from $\Psi=0^0$ and goes with  10$^0$ steps, it is not a surprise that the low-field absorption peak with such a peculiar angular dependence was overlooked in previous studies.

\begin{figure}[ht]
\includegraphics[width=0.45\textwidth]{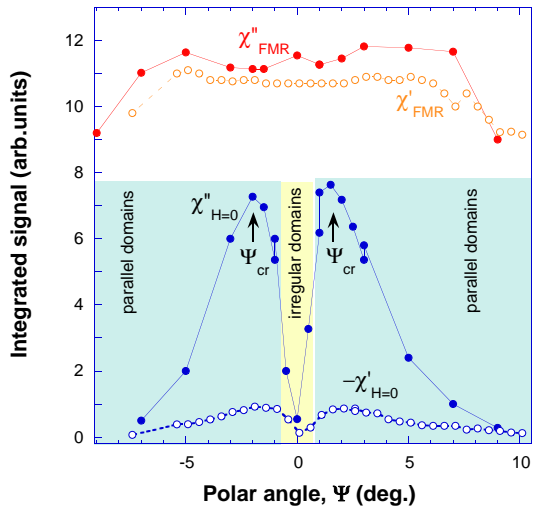}
\caption{Angular dependence of the  microwave susceptibility at zero field, $\chi''_{H=0},\chi'_{H=0}$;  and at the ferromagnetic resonance, $\chi''_{FMR}$,$\chi'_{FMR}$ (peak-to-peak). The  filled circles stay for  absorption, the open circles stay for  dispersion.  The low-field absorption peak  appears only at some deviation from the perpendicular orientation when the parallel domains are expected (gray area). The absorption peak disappears when the deviation from the perpendicular orientation is less than $1^0$ (yellow area) when the domain pattern is, presumably, irregular.
}
\label{fig:angular}
\end{figure}

Figure \ref{fig:angular} shows angular dependence of the maximum low-field absorption and dispersion, $\chi'_{max},\chi''_{max}$, as well as the magnitude of the FMR peak, $\chi'_{FMR},\chi''_{FMR}$. In the narrow angular range  around perpendicular orientation  the latter are nearly constant, while the former  appear at $|\Psi|<8^{0}$, grow towards perpendicular orientation and suddenly disappear in the even more narrow angular range of $|\Psi|< 1^{0}$.

\subsubsection{Field dependence} 
\begin{figure}[ht]
\includegraphics[width=0.4\textwidth]{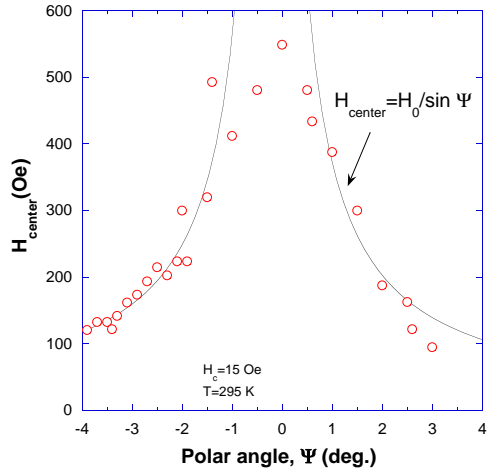}
\caption{The field at which the maximum zero-field absorption appears at $T=295$ K. The solid lines show approximation by the following dependence, $H_{center}=H_{0}/sin\Psi$,  where $\Psi$ is the polar angle and $H_{0}=7.8$ Oe.
} 
\label{fig:center}
\end{figure}

The  low-field absorption  achieves its maximum value $\chi'_{max}$ at small but non-zero field, $H_{center}$, that depends on the  direction of the field sweep (Fig.\ref{fig:der-abs}) and on the field orientation (Fig.\ref{fig:center}).  The angular dependence of this field is satisfactorily accounted by the following function $H_{center}\approx H_{0}/\sin{\Psi}$ where $H_{0}=$ 7.8 Oe which  is close to  the coercive field.  This means that the maximum  absorption is achieved when the in-plane  component of the magnetic field is close to the coercive field that is consistent with  Eq.\ref{eigenfrequency3} (since  $M_{x}\approx M$ at $H_{y}=H_{c}$ then $\omega_{h}$ most closely approaches $\omega$).  

\begin{figure}[ht]
\includegraphics[width=0.45\textwidth]{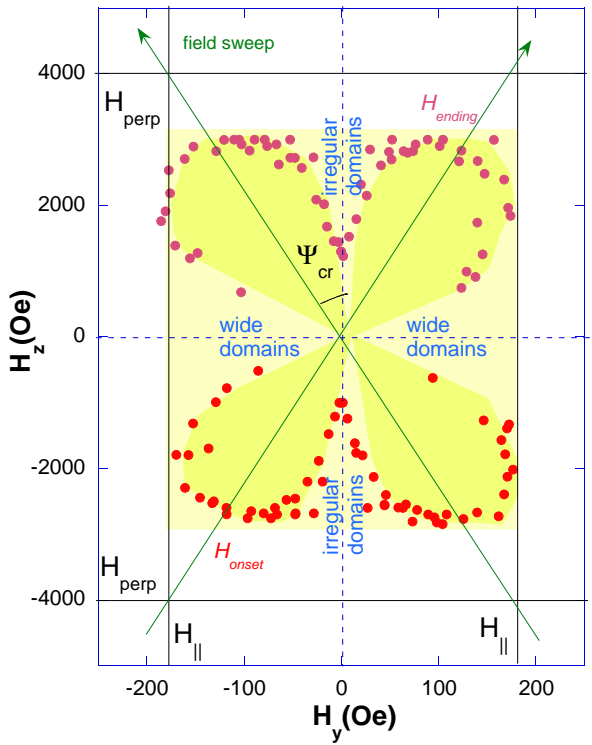}
\caption{The range of magnetic fields where the low-field microwave absorption peak appears. The red and purple circles stand, correspondingly, for the  onset  and the ending fields at which the low-field absorption peak is observed upon the field sweep from -10 kOe to 10 kOe (see Fig.\ref{fig:int-abs1}). Note different horizontal and vertical scales.  $H_{\perp}$ and $H_{\parallel}$ stand for the perpendicular and parallel anisotropy fields. The pale yellow area indicates the field range where domain structure can appear.  The  bright yellow area enclosed by the data points shows the field range where the low-field microwave absorption appears. 
}
\label{fig:map}
\end{figure}

Figure \ref{fig:map} shows the field range where the low-field absorption peak is observed at 295 K. This range has a four-lobed shape with the lobes oriented along the directions $|\Psi|=2.6^{0}$. Remarkably,  this value coincides with the critical angle that was defined in relation to Eq.\ref{split}. Indeed, by substituting into Eq.\ref{split} the perpendicular anisotropy field, $H_{\perp}=$ 4  kOe (measured from the FMR in the perpendicular geometry), and  the parallel anisotropy field, $H_{\parallel}=$ 180 Oe (found from  magnetization measurements) we find $\Psi_{cr}=H_{\parallel}/H_{\perp}=2.6^{0}$. 
The lobes are within the rectangle defined by the conditions: $H_{y}<H_{\parallel}, H_{z}<0.7H_{\perp}$. The former  defines the onset of the multidomain state for the nearly parallel field orientation, while the latter  means that the magnetization lies predominantly in-plane (more precisely, the deviation of the magnetization from the in-plane orientation  does not exceed $45^{0}$). 

\subsubsection{Comparison to the model}
The unusual angular dependence of the low-field absorption peak (Fig.\ref{fig:angular}) and its hysteretic behavior  (Fig.\ref{fig:der-abs}) imply that the peak magnitude strongly depends on the field orientation during magnetic field sweep. How  such dependence on magnetic history can be qualitatively explained by Eqs.\ref{eigenfrequency2},\ref{chi}? The film has out-of-plane hard axis in the $z$-direction and the in-plane hard axes in the $x-$ and $y-$ directions. Figure \ref{fig:map} shows that the low-field absorption appears when the film is in the unsaturated state both with respect to the out-of-plane and  in-plane hard axes. This state is presumably the multidomain state. Indeed, there are quite a few observations of magnetic domains in the LSMO films on STO \cite{Lecoeur,Dho,Murakami,Taniuchi,Ziese,Houwman}. Photoelectron emission spectroscopy \cite{Taniuchi} revealed  3-30 $\mu$m wide parallel domains, magnetic force microscopy showed much smaller, 1 $\mu$m wide domains \cite{Ziese}, or the checkerboard domain pattern with  0.5-0.75 $\mu$m wide domains \cite{Houwman}, and even smaller 0.3 $\mu$m wide domains (in LCMO thin films) \cite{Lloyd}. To explain our results we assume  parallel domains  whose width and orientation depend on the values of  $H_{z}$ and $H_{y}$ at the onset of the domain state, hence the dependence on magnetic history.

We attribute the  absorption peaks shown in Figs.\ref{fig:int-abs}, \ref{fig:int-abs1}  to the tail of the domain mode resonance. The peak magnitude is set by the proximity of the resonant frequency $\omega_{h}$  to the operating frequency $\omega$. Dependence of $\omega_{h}$ on magnetic history is captured by the parameter $N_{y}$ (Eq.\ref{eigenfrequency3}) which is determined by the domain width $w$. The latter is found by minimizing the sum of the energy associated with the stray field of domains and the total energy of domain walls $W_{DW}$ \cite{Hubert}.  The onset of the multidomain state is set by $H_{y}$ while the domain wall energy and, correspondingly, the domain width are set by  $H_{z}$. The latter can strongly affect the energy and structure of the domain walls and even drive the transition from the Bloch to N\'{e}el wall \cite{Hubert,Grishin}. When the field  is lowered, the domain width changes but due to  pinning some memory of the magnetic conditions at the onset of the multidomain state is conserved.  Since $H_{y}$ and $H_{z}$ play very different roles in establishing the domain state, the domain pattern and the frequency of the domain mode resonance (Eq.\ref {eigenfrequency3}) depend on field orientation during field sweep, hence the resulting absorption peak is strongly angular- and field-dependent. We can speak on "angle-tuned" domain mode resonance.  

 When the field is parallel to the film, domain width is large and $\omega_{h}$ is too low to produce appreciable absorption at $\omega/2\pi=$9.4 GHz.  When the field is close to the perpendicular orientation, the domain width is small, $\omega_{h}$ approaches $\omega$ and absorption strongly increases. When the field is too close to the perpendicular orientation there is no preferential domain wall orientation, domain pattern is irregular, and the resonance is not excited. These considerations can be succinctly summarized as follows: $H_{z}\rightarrow W_{DW}\rightarrow w\rightarrow N_{y}\rightarrow\omega_{h}\rightarrow \chi_{max}$.


\subsection{Low-field microwave absorption at low temperatures}

\begin{figure}[ht]
\includegraphics[width=0.45\textwidth]{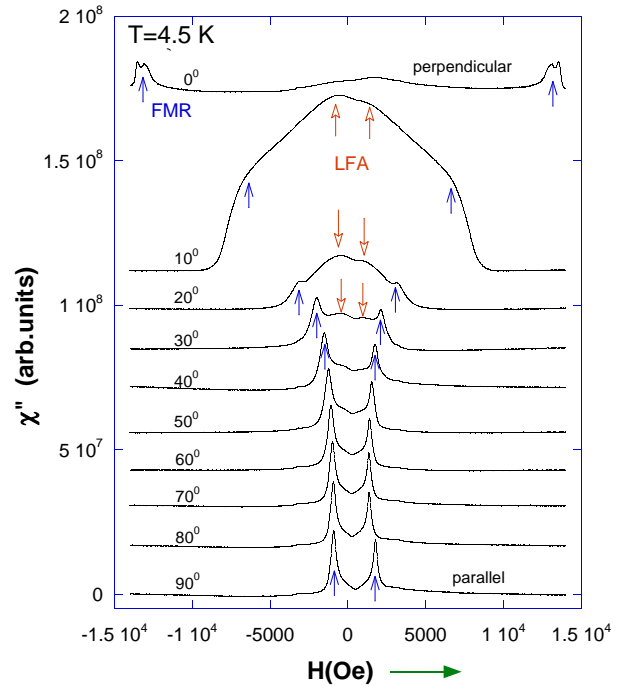}
\caption{Integrated absorption for the sample of the Fig.\ref{fig:int-abs1} at 4.5 K at different  field orientations. The purple arrows indicate  FMR peaks, the red arrows indicate low-field absorption peaks associated with the domain mode resonance. The FMR peaks appear in all orientations while the domain mode resonance appears only in oblique orientation and disappears both in the parallel and perpendicular orientations. Unlike Fig.\ref{fig:int-abs1} where at each field orientation there is a single low-field absorption peak, here there are two overlapping peaks.
} 
\label{fig:zero-4K}
\end{figure}

Table I demonstrates  that the maximum low-field absorption  $\chi''_{max}$ (achieved at different angles for each temperature)  does not disappear at low temperatures and even increases. This stays in sharp contrast with the microwave losses associated with magnetoconductance (magnetoresistance), whose magnitude for our highly conducting films  dramatically decreases at low temperatures. This proves once again that the low-field microwave absorption is unrelated to magnetoresistance. We observe also that $\chi''_{max}$ is on the same order as the FMR absorption peak in the parallel geometry. This is not surprising since according to our model the zero-field absorption peak is nothing else but the FMR absorption peak in the parallel geometry which is shifted to zero field due to dynamic demagnetization factor associated with magnetic domains.
\begin{table}
\caption{\label{tab:table1} Comparison of different contributions to microwave absorption.   (The units are arbitrary but the same for all entries).
}
\begin{ruledtabular}
\begin{tabular}{cccc}
$T$(K)&$\chi''_{H=0}$\footnote{The  maximum low-field microwave absorption (at corresponding angle). }&$\chi''_{FMR}$\footnote{The magnitude of the FMR absorption peak in  parallel field.}&$\chi''_{MR}$\footnote{Nonresonant microwave losses associated with magnetoresistance in parallel field, $\chi''_{MR}=\chi''_{H=10 kOe}-\chi''_{H=0}$.} \\
\hline
295 & 3.3 & 2.5& 31\\
270 & 3 & 1& 12\\
220 & 2.5 & 3.4& 6.2\\
110& 4.3 & 4& 2.8\\
4.5 & 8.8 & 2.6& $<$0.3\\
\end{tabular}
\end{ruledtabular}
\end{table}

The temperature dependence of the low-field absorption is dictated mostly by $\omega_{h}$ and its proximity to the operating frequency $\omega$ (Eq.\ref{eigenfrequency3}). The low-field absorption at 295 K arises from the tail of the domain mode resonance since $\omega_{h}<\omega$.  To estimate $\omega_{h}$ at 4.5 K we note that  at 4.5 K 
$4\pi M=$ 7.4 kG (saturation magnetization of LSMO). We substitute this value and  $d=$ 200 nm, and $w=$0.5 $\mu$m into Eq.\ref{eigenfrequency3} and find $\omega_{h}/2\pi=$12.2 GHz, in such a way that $\omega_{h}>\omega$.  This means that the maximum absorption is achieved at resonant field that satisfies the condition $\omega_{h}=\omega$. Upon field sweep across the zero this condition is met twice- once for negative and once for positive field. 

Figure \ref{fig:zero-4K} shows the low-temperature microwave absorption spectra at different angles. In contrast to Fig.\ref{fig:int-abs1} there are two low-field absorption peaks: one at negative  and one at positive field. 
There are other differences with respect to Fig.\ref{fig:int-abs1} as well. The baseline in Fig.\ref{fig:zero-4K} is flat indicating negligible contribution  from magnetoresistance, in such a way that in the parallel orientation the absorption spectrum exhibits only the FMR peak.   When the field is rotated away from the parallel orientation there appears a low-field absorption peak that becomes a dominant feature at $\Psi= 5^{0}-20^{0}$. This peak  disappears again when the deviation from the perpendicular orientation is less than $1^0$. 

\section{Conclusions} In thin La$_{0.7}$Sr$_{0.3}$MnO$_{3}$ films with easy-plane anisotropy  there is a special kind of the domain mode resonance that is absent in the parallel and perpendicular field orientation and appears only in the oblique magnetic field. It can be "angle-tuned" by varying the field orientation with respect to the film. The resonance in oblique field should not be unique to manganites and it can be a source of pronounced low-field microwave absorption in other magnetic materials as well.

\section{Acknowledgments} We are grateful to S. Mercone, S.V. Kapel'nitskii, and K.N. Rozanov for stimulating discussions and constructive comments.


\begin{thebibliography}{}
\bibitem{Haddon}R. C. Haddon, S. H. Glarum, S. V. Chichester, A. P. Ramirez, and N. M. Zimmerman, Phys. Rev. B \textbf{43}, 2642 (1991). 
\bibitem{Bohandy}B.F. Kim, J. Bohandy, K. Moorjani, and F.J. Adrian, J. Appl. Phys. \textbf{63}, 2029 (1988).
\bibitem{Owens} F.J. Owens, J. Phys. Chem. Sol. \textbf {58}, 1311 (1997);\emph{ibid.} \textbf {66}, 793 (2005).
\bibitem{Veinger}A. I. Veinger, A. G. Zabrodskii, T. V. Tisnek, Supercond. Sci. Technol. \textbf{8}, 368 (1995).
\bibitem{Dionne} G.F. Dionne, IEEE Trans. Magn. \textbf{39}, 3121  (2003).
\bibitem{Valenzuela}R. Valenzuela, R. Zamorano, G. Alvarez, M.P. Gutierrez, H. Montiel, J. Non-Crystalline Solids \textbf{353}, 768 (2007).
\bibitem{Montiel}H. Montiel, G. Alvarez, R. Zamorano, R. Valenzuela, J. Non-Crystalline Solids \textbf{353}, 908 (2007)
\bibitem{Stanescu} D. Stanescu, P. Xavier, J. Richard, and C. Dubourdieu, J. Appl. Phys. \textbf{99}, 073707 (2006).

\bibitem{Bahlmann}N. Bahlmann, R. Gerhardt, H. Doetsch, J. Magn. Magn. Materials, \textbf{16l}, 22 (1996).
\bibitem{Synogach} V.T. Synogach and H. Dotsch, Phys. Rev. B \textbf{54}, 15266 (1996). 
\bibitem{Vukadinovic2000} N. Vukadinovic, J. Ben Youssef, H. Le Gall, J. Ostorero, Phys.Rev. B. \textbf{62}, 9021 (2000).

\bibitem{Prinz} G.A. Prinz, G.T. Rado, J.J. Krebs, J. Appl. Phys. \textbf{53}, 2089 (1982).
\bibitem{Baberschke} F. Gerhardter, Yi Li, K. Baberschke, Phys. Rev. B \textbf{47}, 11204 (1993).
\bibitem{Lofland}  S.E. Lofland, V. Ray, P.H.  Kim , S.M. Bhagat, K. Ghosh, R.L. Greene, S.G. Karabashev, D.A. Shulyatev, A.A. Arsenov, Y. Mukovskii, J. Phys. Cond. Matter, \textbf{9}, L633 (1997).
\bibitem{Lee} S. J. Lee, C. C. Tsai, H. Cho, M. Seo, T. Eom, W. Nam, Y. P. Lee,and J. B. Ketterson, J. Appl. Phys. \textbf{106}, 063922 (2009).
\bibitem{Rivoire} M. Rivoire and G. Suran, J. Appl. Phys. \textbf{78}, 1899 (1995).

\bibitem{Artman} J.O. Artman, Phys. Rev. B \textbf{105}, 74 (1957).
\bibitem{Hasty} T.E. Hasty, J. Appl. Phys. \textbf{35}, 1434 (1964).
\bibitem{Salansky}N. M. Salansky, B.P. Khrustalev, Czech. J. Phys. \textbf{21}, 419 (1971).
\bibitem{Buznikov} N.A. Buznikov, K.N. Rozanov, J. Magn. Magn. Materials \textbf{285}, 314 (2005).
\bibitem{Layadi}A. Layadi, F. W. Ciarallo, J. 0. Artman, IEEE Trans. Magn., \textbf{23}, (1987). 
\bibitem{Vukadinovic1995} N. Vukadinovic, J. Ben Youssef, H. Le Gall, J. Magn. Magn. Mat. \textbf{150}, 213 (1995).
\bibitem{Ebels} U. Ebels, P.E. Wigen, K. Ounadjela, J.Magn.Magn Materials \textbf{177}, 1239 (1998).
\bibitem{Bi}S.-Y. Bi, J. Appl. Phys. \textbf{67}, 3179 (1990).
\bibitem{Acher}O. Acher, C. Boscher, B. Brule, G. Perrin, N. Vukadinovic, G. Suran, H. Joisten, J. Appl. Phys. \textbf{81}, 4057 (1997).
\bibitem{Zakhidov} A.A. Zakhidov, I.I. Khairullin, V.Y. Sokolov,
R.H. Baughman, Z. Iqbal, M. Maxfield, B.L. Ramakrishna, Synthetic Metals \textbf{41-43}, 3717 (1991).
\bibitem{Dubowik}J. Dubowik, F. Stobiecki, P.Y. Goscianska, Czech. J. Physics  \textbf{52}, 227 (2002).

\bibitem{Fukui}K. Fukui, K. Uno, H. Ohya-Nishiguchi, and H. Kamada, J. Appl. Phys. \textbf{83}, 2158 (1998).
\bibitem{Krebs} J. J. Krebs, P. Lubitz, A. Chaiken, and G. A. Prinz, J.Appl.Phys. \textbf{69}, 4795 (1991).
\bibitem{Vittoria} F.J. Rachford, P. Lubitz, C. Vittoria, J. Appl. Phys. \textbf{53}, 8940 (1982). 
\bibitem{Belyaev} B.A. Belyaev, A.V. Izotov, A.L. Leksikov, IEEE Sensors Journal, \textbf{5}, 260 (2005).
\bibitem{Demidov}V. V. Demidov, I. V. Borisenko, A. A. Klimov, G. A. Ovsyannikov, A. M. Petrzhik, and S. A. Nikitov, J.  Exp.  Theor. Phys.  \textbf{112},  825 (2011). 

\bibitem{Salamon}M.B. Salamon, M. Jaime, Rev. Mod. Phys. \textbf{73}, 583 (2001).
\bibitem{Budak} S. Budak, M. Ozdemir, B. Aktas, Physica B \textbf{339}, 45 (2003).

\bibitem{Lyfar}D. L. Lyfar, S. M. Ryabchenko, V. N. Krivoruchko, S. I. Khartsev, and A. M. Grishin, Phys. Rev. B \textbf{69}, 100409(R) (2004).
\bibitem{Golos} M. Golosovsky, P. Monod, P. K. Muduli, R. C. Budhani, L. Mechin, and P. Perna, Phys. Rev. B \textbf{76}, 184414 (2007).


\bibitem{Musa}M.Golosovsky, M.Abu-Teir,D. Davidov, O.Arnache, P.Monod, N.Bontemps, R.C.Budhani, J. Appl. Phys. \textbf{98}, 084902 (2005).
\bibitem{Aswal} D.K. Aswal, A. Singh, R.M. Kadam, M.K. Bhide, A.G. Page, S. Bhattacharya, S.K. Gupta, J.V. Yakhmi, V.C. Sahni, Materials Letters \textbf{59}, 728 (2005).
 


\bibitem{Budhani}K. Senapati and R. C. Budhani, Phys. Rev. B \textbf{71}, 224507 (2005).
\bibitem{Mechin}F. Yang, L. Mechin, J. M. Routoure, B. Guillet, and R. A. Chakalov, J. Appl. Phys. \textbf{99}, 024903 (2006).

\bibitem{Hubert} A. Hubert, R. Schaefer, \emph{Magnetic Domains. The Analysis of Magnetic Microstructures}, Springer, Berlin-Heidelberg-New York, 1998. p.238.

\bibitem{Grishin} S. A. Manuilov, A. M. Grishin, and M. Munakata, J. Appl. Phys. \textbf{109}, 083926 (2011). 

\bibitem{Lecoeur} P. Lecoeur, Pl. Touilloud, Gang Xiao, A. Gupta, G.Q. Gong, X.W. Li, J. Appl. Phys. \textbf{82}, 3934 (1997).

\bibitem{Dho}J. Dho, Y. N. Kim, Y. S. Hwang, J. C. Kim, and N. H. Hur, Appl. Phys. Lett. \textbf{82}, 1434 (2003).

\bibitem{Murakami}Y. Murakami, H. Kasai, J. J. Kim, S. Mamishin, D. Shindo, S. Mori, and A. Tonomura, Nature Nanotechnology \textbf{5}, 37  (2010). 

\bibitem{Taniuchi}T. Taniuchi et al. Appl. Phys. Lett. \textbf{89}, 112505 (2006).





\bibitem{Ziese} M. Ziese, Phys. stat. sol. (b) \textbf{243}, 1383 (2006).

\bibitem{Houwman}E. P. Houwman, G. Maris, G. M. De Luca, N. Niermann, G. Rijnders, D. H. A. Blank, and S. Speller, Phys. Rev. B \textbf{77}, 184412 (2008).





\bibitem{Lloyd}S. J. Lloyd, N. D. Mathur, J. C. Loudon, and P. A. Midgley, Phys. Rev. B \textbf{64}, 172407 (2001). 

\end{thebibliography}
\end{document}